\documentstyle[aps,twocolumn,epsfig]{revtex}
\newcommand{\beq}{\begin{equation}}
\newcommand{\eeq}{\end{equation}}
\newcommand{\bea}{\begin{eqnarray}}
\newcommand{\eea}{\end{eqnarray}}
\newcommand{\ba}{\begin{array}}
\newcommand{\ea}{\end{array}}
\newcommand{\bc}{\begin{center}}
\newcommand{\ec}{\end{center}}

\newcommand{\etal}{{\it et al.}}
\newcommand{\bml}{\begin{mathletters}}
\newcommand{\eml}{\end{mathletters}}
\newcommand{\commentout}[1]{{}}
\newcommand{\k}{{\bf k}}

\newcommand{\q}{{\bf q}}

\newcommand{\eq}[1]{(\ref{#1})}
\newcommand{\comment}[1]{{}}
\begin{document}
\draft
\preprint{}
\wideabs
{
\title{Rate limit for photoassociation of a Bose-Einstein condensate}
\author{Juha Javanainen$^\dagger$ and Matt Mackie$^\ddagger$}
\address{$^\dagger$Department of Physics, University of Connecticut, 
Storrs,
Connecticut
06269-3046
\\
$^\ddagger$Helsinki Institute of Physics, PL 64, FIN-00014 Helsingin
yliopisto,
Finland}
%\author{Matt Mackie}
%\address{}
\date{\today}
\maketitle
\begin{abstract}
We simulate numerically the photodissociation of molecules into
noncondensate
atom pairs that accompanies photoassociation of an atomic
Bose-Einstein condensate into a molecular condensate.  Such rogue
photodissociation sets a limit on the achievable rate of
photoassociation. Given the atom density
$\rho$ and mass
$m$, the limit is approximately $6\hbar\rho^{2/3}/m$. At low 
temperatures
this is a more stringent restriction than the unitary limit of
scattering theory.
\end{abstract}
}
\narrowtext
Photoassociation of a Bose-Einstein condensate~\cite{DRU98,JAV99} has
been predicted to feature examples of quantum coherence, such
as Rabi
oscillations between atomic and molecular
condensates~\cite{JAV99,HEI00}, rapid~\cite{JAV99} and 
Raman~\cite{MAC00}
adiabatic passage from an atomic condensate to a molecular
condensate, and a superposition of macroscopic numbers of atoms and
molecules~\cite{CAL01}. The Feshbach resonance in a magnetic field is
mathematically equivalent  to
photoassociation, and should give rise to analogous
phenomena~\cite{TIM98}.
Spontaneous-emission decay of the photoassociated molecules causes
experimental
problems~\cite{KOS00}, and  so far experiments on photoassociation of a
Bose-Einstein condensate~\cite{WYN00,GER00} have not demonstrated
coherent
optical transients. Nonetheless, the prospect of a molecular condensate
continues to motivate further work.

Many of the past theoretical studies of coherent photoassociation have
taken into account only the atomic and molecular condensates. Notably
ignored is the photodissociation of a molecule into a
pair of atoms, neither of which belongs to the atomic condensate. We
have
investigated such ``rogue'' photodissociation~\cite{JAV99,KOS00} using
essentially dimensional arguments, and suggested that there is a
shortest
possible time scale and a corresponding fastest possible Rabi frequency
for
coherent photoassociation. Given the laser intensity
$I$, the Rabi frequency
$\Omega$ is proportionally to the laser field strength,
$\Omega\propto\sqrt I$, whereas the rate of rogue photodissociation
scales as
$\Gamma\propto I$. The equality $\Omega=\Gamma$ occurs when both of
these
frequencies are approximately equal to
\begin{equation}
{1\over\tau_\rho}=\omega_\rho = {\hbar\rho^{2/3}\over 2\mu}\,.
\label{SCALES}
\end{equation}
Here $\rho=\rho_A+2\rho_M$  is an invariant density derived from the
densities
of the atoms and the molecules, and $\mu$ is the reduced mass of a pair
of
atoms.  In an attempt to speed up coherent photoassociation beyond
this limit by
increasing the laser intensity, damping due to rogue photodissociation
overtakes coherent photoassociation~\cite{JAV99,KOS00}. There have also
been
numerical case studies on the effects of noncondensate modes on
photoassociation\cite{GOR01,HOL01}, but these do not spell out time
or frequency
scales comparable to Eq.~(\ref{SCALES}).

In the present paper we formulate a minimal yet realistic, microscopic 
model
for  rogue
photodissociation, and solve it numerically. The results demonstrate
that
there is a maximum achievable photoassociation rate, of the
order $6\omega_\rho$. The existence of the rate limit is at variance
with both
the wave picture of condensate atoms, and the standard scattering
theory.

We take a plane-wave laser field with photon momentum $\q$ to drive
photoassociation and photodissociation.  Initially there is only
a condensate of $N$ zero-momentum atoms present. By momentum
conservation,
only molecules with momentum $\q$ will be generated in the primary
photoassociation. In photodissociation such molecules break up into a
pair of
atoms with opposite momenta. Processes in which both atoms return to
the
condensate are the subject of our earlier work on coherent
photoassociation~\cite{JAV99,KOS00}, whereas here the focus is on
processes in
which the molecules break up into a pair of noncondensate atoms.

Our description starts from our generic Hamiltonian
for photoassociation and dissociation~\cite{JAV99,KOS00}. We write
the Heisenberg
equations of motion for the boson operators $a_\k$ ($b_\k$)
characterizing
atoms (molecules) with momentum $\k$, and implement two main
approximations.
First, in the classical approximation analogous to the use of the
Gross-Pitaevskii equation, we treat the atomic and molecular condensate
operators $a\equiv a_0$ and $b=b_\q$ as $c$-numbers. Second, we take
into
account all pairs of noncondensate atoms with opposite momenta, but no
further
photoassociation of noncondensate atoms to any molecular states except
back to
the molecular condensate. It then turns out that the equations of
motion for
the $c$-number atomic amplitude
$\alpha =  a/\sqrt{N}$, molecular amplitude $\beta= \sqrt{2/N}\, b$,
and noncondensate atom pair amplitudes $c_\k = \langle a_\k a_{-\k}
\rangle$
close. As usual, we discuss
$s$-wave photoassociation and dissociation, so that we may uniquely
label
the two-atom modes with their energy $\hbar\epsilon$. The equations of
motion may be cast in the form
\bml
\bea
\dot{\alpha} &=& i{\Omega\over\sqrt{2}}\,\alpha^*\beta,\label{ALEQ}\\
\dot{\beta} &=& i\delta\beta + i{\Omega^*\over\sqrt{2}}\,\alpha^2+i\int
d\epsilon\,\xi(\epsilon)\,c_\epsilon,\label{BLEQ}\\
\dot{c}_\epsilon &=& -i\epsilon\,c_\epsilon + i
\xi^*(\epsilon)\beta\,\label{CEQ}.
\eea
\label{EQM1}
\eml

To explain the notation, we first consider the model in the absence of
the
coupling between condensate atoms and molecules $\propto\Omega$. The
quantity
$\delta$ is the detuning of the laser above the
photodissociation threshold of the molecules (corrected for photon
recoil
energy~\cite{JAV99,KOS00}), and
$\xi(\epsilon)$ is the coupling strength for photodissociation of a
molecule
to an atom pair with energy
$\hbar\epsilon$. Using the Fermi golden rule, we find the
photodissociation rate for molecules $\Gamma(\delta) = 2\pi
|\xi(\delta)|^2$.
In our modeling we turn the tables, and assert the equality
\beq
\xi(\epsilon) =
\sqrt{\Gamma(\epsilon)/2\pi}\,
\eeq
relating the matrix element $\xi(\epsilon)$ and the on-shell
photodissociation
rate
$\Gamma(\epsilon)$  to an atom pair with energy $\hbar\epsilon$.
Second, we know
from our theory of coherent photoassociation~\cite{JAV99,KOS00} that
the Rabi
frequency
$\Omega$ is related to the rate of photodissociation by
\beq
\Omega =
\lim_{\epsilon\rightarrow0}
\sqrt{{\sqrt2\pi\hbar^{3/2}\rho\over\mu^{3/2}
}\,{\Gamma(\epsilon)\over \sqrt{\epsilon}}}\label{OMEQ}\,.
\label{OME}
\eeq
Third, in order for Eq.~\eq{OME} to make sense, for low energies
$\Gamma(\epsilon)\propto\sqrt\epsilon$ must hold true. This, in fact,
is known
as the Wigner threshold law. In our modeling we take the threshold law
to hold
for all relevant energies. Finally, for notational convenience, we
write the
continuum amplitude in terms of a dimensionless quantity $C(x) \equiv
c_{x\Omega}\Omega^{-1/2}$. Our final equations of motion for the
amplitudes
read
\bml
\bea
\dot{\alpha} &=& i{\Omega\over\sqrt{2}}\,\alpha^*\beta,\label{ALEQ2}\\
\dot{\beta} &=& i\delta\beta + i{\Omega\over\sqrt{2}}\,\alpha^2+i
\gamma^{1/2}\Omega\int
dx\,\sqrt{x}\,
C(x),\label{BLEQ2}\\
\dot{C}(x) &=& -i\Omega x\,C(x) + i \Omega
\gamma^{1/2} \sqrt{x}\,\beta\,\label{CEQ2}.
\eea
\label{EQM2}
\eml
The parameter
\beq
\gamma = {\Gamma(\Omega)\over2\pi\Omega} =
{1\over8\pi^2}\,\left({\Omega\over\omega_\rho }\right)^{3/2}
\eeq
characterizes the relative importance of rogue
photodissociation and coherent photoassociation.

In Eqs.~\eq{EQM2} we could readily absorb the frequency $\Omega$ into
the unit
of time, but we will not do so as this would cause contorted
discussions below.
Coupling to the noncondensate modes introduces the frequency
$\omega_\rho$.

We study Eqs.~\eq{EQM2} numerically, by discretizing the dissociation
continuum.
We also cut off the continuum at a maximum frequency $\epsilon_M$.
Basically, the vector made of the amplitudes
$\alpha$, $\beta$ and $C(x_n)$, $\psi$, satisfies a  Schr\"odinger
equation of
the form $i\dot\psi = H(\psi)\psi$. As Eqs.~\eq{EQM2} are nonlinear,
the
Hamiltonian $H$ depends on the vector $\psi$ itself. However,
if this dependence is ignored, the resulting linear Schr\"odinger
equation
may be solved using the norm conserving algorithm described in
Ref.~\cite{JAV99b}. Moreover, the matrix of the Hamiltonian is such
that the run
time scales linearly with the number of the continuum states
$n_c$. We take care of the nonlinearity with a predictor-corrector
method. To
execute a time step $t\rightarrow t+h$, we first integrate the
Schr\"odinger
equation for the constant Hamiltonian $H[\psi(t)]$ to obtain the
``predicted''
wave function
$\tilde\psi(t+h)$, and then ``correct'' by integrating the
Schr\"odinger
equation anew using the Hamiltonian ${1\over2}\{H[\psi(t)] +
H[\tilde\psi(t+h)]\}$. The algorithm is norm conserving, the error
for integrating over a fixed time interval scales with the step size
at least as ${\cal O}(h^2)$ (empirically often better), and the run
time
is linear in the number of continuum states. Typically we discretize
the
continuum in $n_c=10^4$ steps with the maximum frequency
$\epsilon_M=100\,\Omega$, but when needed for convergence we do not
hesitate
to resort to, say, $n_c=10^6$ and $\epsilon_M=10^3\,\Omega$.

Coupling to the dissociation continuum also shifts the molecular
level~\cite{MAC99,GER01}.   We
correct for the shift by adding to the detuning the Stark shift
calculated using
perturbation theory for on-threshold photodissociation, namely,
$\Gamma(\epsilon_M)/\pi$. The shift may be very large; in fact, if we
did not
truncate the continuum, it would be infinite. This is our version of
the
renormalization problem discussed in Ref.~\cite{HOL01}. In the limit of
a large
dissociation rate, perturbation theory becomes unreliable.
Correspondingly,
it gets difficult to tell exactly what the detuning is with respect to
the
photodissociation threshold. This problem is not unique to our model,
but would emerge in real experiments as well.

For demonstration, Fig.~1 gives the quantities $P_A = |\alpha|^2$ and
$P_M=|\beta|^2$, the probabilities that an arbitrary atom in the system
is
part of the atomic or the molecular condensate, as a function of time,
given
that the system started as an atomic condensate. In Fig.~1(a) we set
$\gamma=0$, and thereby disregard rogue photodissociation. For this
(renormalized) detuning
$\delta=\Omega$, the conversion between atoms (dotted line) and
molecules (dashed
line) oscillates in such a way that  half of the atomic
condensate is periodically converted into a molecular condensate. In
Fig.~1(b)
we add rogue photodissociation, setting
$\Gamma(\Omega)=\Omega$ and hence $\gamma=1/(2\pi)$. After a few damped
oscillations, the probabilities settle down.
The total probability that an atom is part of either condensate (solid
line)
drops to 0.8, indicating that in the process 20~\% of the atoms are
lost
to rogue photodissociation. Although the couplings between the
condensates and
from the molecular condensate to the noncondensate atoms remain, the
atomic and
the molecular condensates lock up in a superposition immune to rogue
photodissociation. It is known from the past work on decay of a
discrete
state to a shaped continuum (e.g., photodissociation of a negative
ion~\cite{RZA82}) that, given a threshold to the continuum, some
population may stay permanently trapped in the bound state. Curiously,
the same
applies here to a superposition of atomic and molecular amplitudes,
even though
the condensate system is nonlinear and the superposition principle does
not
hold.

Our main subject is limitations on conversion from atoms to molecules
due to
rogue photodissociation. In the traditional perturbation theory one
would study
the photoassociation rate as a function of the problem parameters.
However, as
is obvious from Fig.~1, atom-molecule conversion is not governed by a
simple exponential time dependence. A clear-cut photoassociation rate
generally does not exist. Because of level shifts, it may also be
difficult to know the true value of the detuning. To circumvent these
problems, we have developed a protocol that mimics conceivable
experiments. We first fix an interaction time,
call it
$\tau$, once and for all. Our basic numerical operation is to calculate
the loss of condensate atoms as a function  of the
Rabi frequency $\Omega$, rogue photodissociation amplitude $\gamma$,
and
detuning $\delta$, $1-P_A(\Omega,\gamma,\delta; \tau)$, starting from a
pure atomic condensate.  We vary
the
detuning
$\delta$, and find what we interpret as the maximal photoassociation
probability
with the other parameters fixed, $1-P_A(\Omega,\gamma;\tau)$. The 
optimal
rate  for
photoassociation is then defined as
$R(\Omega,\gamma;\tau)=[1-P_A(\Omega,\gamma;\tau)]/\tau$.

To get a quantitative handle on the variation of the effective rate $R$
we
divide it by $\omega_\rho$, and plot in Fig.~2 the result as a function
of the
variable $1/(\tau\omega_\rho)$ for four different Rabi
frequencies, $\Omega =  1/\tau$ (dotted line), $2/\tau$ (dash-dotted
line), $4/\tau$ (dashed line), and $8/\tau$ (solid line). It appears
that there is a maximum  rate for photoassociation of the order of
$6\,\omega_\rho$.

We emphasize once more that our maximal photoassociation rate measures
how many
atoms one can remove in a fixed interaction time. There really does not
have to
be a rate at all. We demonstrate this by plotting in
Fig.~\ref{TIMEDEP} the time
dependences of the molecular probability $P_M$ (dotted line), the
probability for condensate atoms $P_A$ (dashed line), and the total
condensate
probability
$P_A+P_M$ (solid line) as a function of time. The fixed parameters
$\Omega =
4/\tau$,
$\delta=-9.27\,\Omega$, and
$\gamma=1.27$ are chosen in such a way that this is the time evolution
used to
obtain the maximum of the curve labeled $\Omega=4/\tau$ in
Fig.~\ref{MAXRATE}.
The laser is tuned well {\em below\/} the photoassociation threshold.
There is
very little molecular probability, so that the molecules are a virtual
intermediate state. Instead, we have a slowly damped two-photon
flopping
between the atomic condensate and noncondensate atom pairs. It is
doubtful that such flopping survives the various sources of
incoherence left out
from our model, such as photoassociation of noncondensate atoms to
noncondensate molecules or collisions of noncondensate atoms. But we
surmise
that added loss mechanisms further hinder photoassociation, and our
upper limit
for the rate $6\,\omega_\rho$ would probably hold in a more complete
model.

So far we have discussed the case when the molecular state is not
damped by
spontaneous emission. Our modeling is most appropriate for two-color
photoassociation in the Raman configuration. However, we have also
carried out
realistic simulations of one-color photoassociation, in which the
spontaneous
decay rate of the photoassociated molecules is the largest frequency
parameter
in the problem. Photoassociation once more does not proceed
exponentially and
there is some ambiguity in the definition of the rate, but the
conclusion is
that the maximum achievable rate for photoassociation very nearly
equals
$\omega_\rho$~\cite{JAV01}.

If one is to think of atoms as particles, there must be a limit to
photoassociation rate. In order to photoassociate, two atoms  have to
drift close to one another. This process takes the longer, the more
dilute the gas is; and it cannot be sped up arbitrarily by increasing 
the
light intensity. Beyond a certain intensity, the rate will no longer
increase with increasing intensity. In fact, for a quantum mechanical,
noninteracting,  zero-temperature condensate, one can construct (up to 
a
numerical factor) only one characteristic frequency, $\omega_\rho$. Not
surprisingly,  (up to a  numerical factor) the maximum photoassociation
rate equals $\omega_\rho$

On the other hand, within the two-mode approximation, a gas can be
photoassociated arbitrarily fast; just set $\gamma=0$ in Eqs.~\eq{EQM2}
and let $\Omega\rightarrow\infty$ by increasing the laser intensity. In
the  two-mode approximation the  atoms are waves, simultaneously
everywhere in the  atomic condensate. The existence of a maximum to the
photoassociation rate  thus is a question of particle versus wave 
nature
of the atoms.

Interestingly, in scattering theory the argument goes in the
opposite direction. Because of the wave nature of particles, the
scattering cross section may be much larger than the geometric size of
the colliding objects. The maximum quantum mechanical cross section for
photoassociation, the unitary limit, gives the same rate limit as rogue
photodissociation when the temperature is of the order of the critical
temperature for Bose-Einstein condensation. However, at {\em lower\/}
temperatures the newly found {\em particle\/} nature dominates, and the
rate limit due to rogue photodissociation is more stringent than the
unitary limit. There is no rogue-photodissociation limit in the
usual resonance scattering analysis of photoassociation~\cite{NAP94}, 
and
an experimental observation would constitute a dramatic failure of
scattering theory.

The two-atom position correlation function straddles particle and wave
pictures. Photoassociation evidently depletes two-body correlations at
short
distances, and the motion of the atoms has to replete them. Though we
have not
been able to make an explicit connection, we surmise that
Ref.~\cite{HOL01}
focusing on atom-atom correlations boils down essentially to the same
physics
as our rogue photodissociation. In Ref.~\cite{GOR01}
the Gross-Pitaevskii equations were actually integrated numerically in
the momentum representation. The phenomenon thus found seems to be
what we call modulational instability~\cite{JAV99a}, not rogue
photodissociation.

In sum, we have proposed and numerically solved a  minimal but 
realistic model
for the photodissociation of molecules to noncondensate atoms that 
invariably
accompanies coherent photoassociation of a Bose-Einstein condensate.
The
results support the notion that, as the atoms need time to collide,
there is a
maximum achievable rate to photoassociation of a condensate. What is
dimensionally obvious but might be physically surprising is that the
maximum
rate in a noninteracting, zero-temperature condensate should be of the
order of
the only frequency parameter that can be constructed out of the density
of the
condensate. Apart from the fact that rogue photodissociation sets
experimental
constraints on coherent photoassociation, experiments would also be
interesting
because they bear on the wave versus particle nature of atoms, and on
the
limits of validity of scattering theory.

This work is supported by  NSF Grants PHY-9801888 and PHY-0097974, by
NASA Grant
NAG8-1428, and was carried out in part at the Aspen Center for Physics.
The
argument that there could be a density-imposed limit on
photoassociation rate
was first made to the authors by Wolfgang Ketterle.

\comment{
TEMPORARY DOCUMENTATION OF NOTATIONAL DIFFERENCES BETWEEN CODE AND THE
PRESENT
PAPER.

In the code, the rate of spontaneous emission comes out as
$\xi^2\sqrt{\epsilon}$. Moreover, in the optimization parts we write
$\xi =
\sqrt{x}\Omega$. This leads to the following identifications between
$x$ and
the parameters in this paper:
\beq
\gamma = {x \Omega^{3/2}\over 2\pi},\quad
\omega_\rho=\left({1\over4\pi x}\right)^{2/3}.
\eeq
}

\begin{figure}
\centering
\epsfig{file=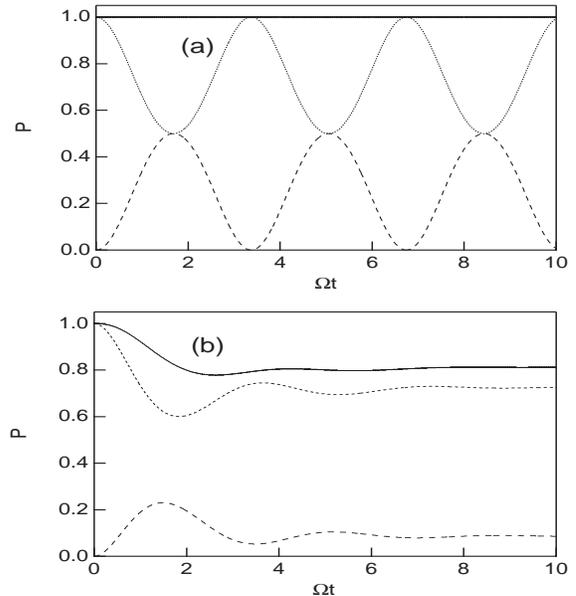,width=8cm,height=8cm}
\caption{Atomic ($P_A$, dotted line) and molecular ($P_M$, dashed
line) probabilities, as well as the total probability that an atom is
part of
either condensate ($P_A+P_M$, solid line) as a function of time. In
panel (a)
there is no rogue photodissociation, $\gamma=0$, while in panel (b) we
set
$\gamma=1/(2\pi)$. The (renormalized) detuning is fixed at
$\delta=\Omega$.}
\label{FLOPPING}
\end{figure}
\begin{figure}
\centering
\epsfig{file=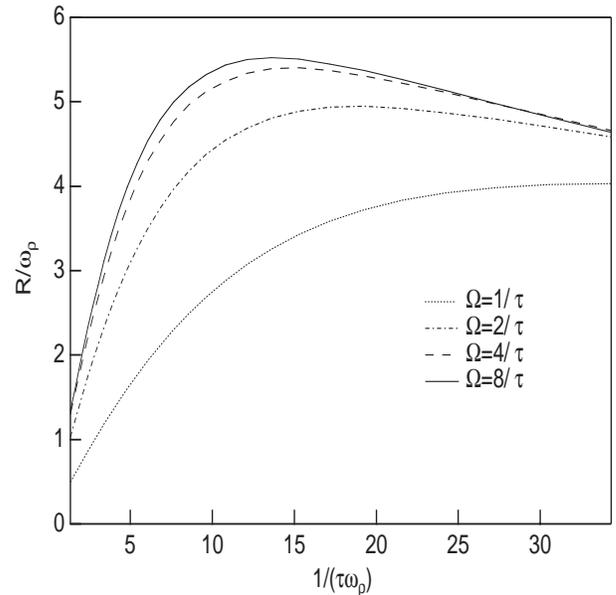,width=8cm,height=8cm}
\caption{Maximum rate of photoassociation $R$ for a fixed interaction
time
$\tau$ scaled to the density dependent frequency parameter
$\omega_\rho$ as a
function of $1/(\tau\omega_\rho$). The lines are for various
Rabi frequencies $\Omega$, as indicated in the legend. Physically, in
order to
follow a line of fixed Rabi frequency toward increasing
$1/(\tau\omega_\rho)$,
one must at the same time increase the laser intensity and decrease the
condensate density, so that rogue photodissociation becomes more
prominent.}
\label{MAXRATE}
\end{figure}
\begin{figure}
\centering
\epsfig{file=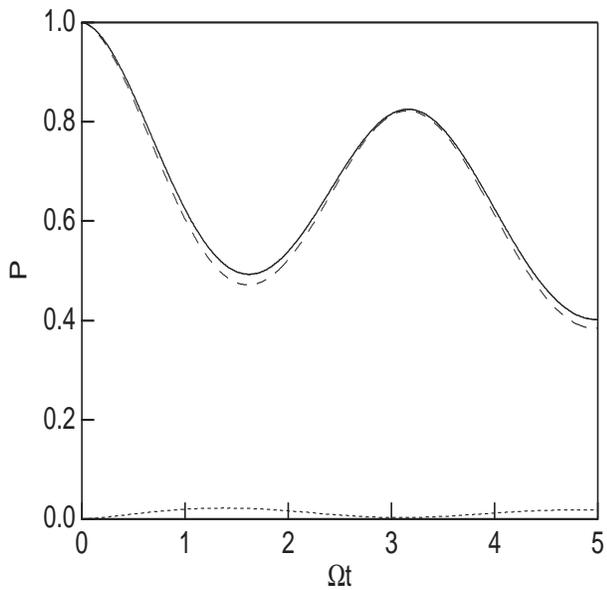,width=8cm,height=8cm}
\caption{
Time
dependences of the three probabilities $P_M$ (dotted line), $P_A$
(dashed
line), and total condensate probability $P_A+P_M$ (solid line) as a
function of
time. The fixed parameters are $\Omega = 4/\tau$,
$\delta=-9.27\,\Omega$,
$\gamma-1.27$. A true photoassociation rate does not exist for these
parameters.}
\label{TIMEDEP}
\end{figure}
\end{document}